\documentclass{eas}
\usepackage{graphicx,amssymb}
\usepackage{natbib,  aas_macros, cases}
\DeclareMathSymbol{\varOmega}{\mathord}{letters}{"0A}
\DeclareMathSymbol{\varSigma}{\mathord}{letters}{"06}

\newcommand{\ts}{t_{\rm stop}}
\newcommand{\taus}{\tau_{\rm s}}

\newcommand{\vc}[1]{\mbox{\boldmath{$#1$}}}

\TitreGlobal{Physics and Astrophysics of Planetary Systems, Les Houches 2008}
\begin{document}

\title{From Grains to Planetesimals}
\runningtitle{Planetesimal Formation}
\thanks{Thanks to Theirry Montmerle and everyone at the Laboratoire d'Astrophysique
de l'Observatoire de Grenoble for a thoroughly enoyable and rewarding school.}
\author{Andrew N. Youdin}\address{C.I.T.A., University of Toronto, 60 St. George St., Toronto ON M5S 3H8 CANADA}
\begin{abstract}
This pedagogical review covers an unsolved problem in the theory of protoplanetary disks: the growth of dust grains into planetesimals, solids at least a kilometer in size.  I  summarize timescale constraints imposed on  planetesimal formation by circumstellar disk observations, analysis of meteorites, and aerodynamic radial migration.  The  infall of $\lesssim$ meter-sized solids in a hundred years is the most stringent constraint.  I review proposed mechanisms for planetesimal formation.  Collisional coagulation models are informed by laboratory studies of microgravity collisions. The gravitational collapse (or Safronov-Goldreich-Ward) hypothesis involves detailed study of the interaction between solid particles and turbulent gas.  I cover the basics of aerodynamic drag in protoplanetary disks, including radial drift and vertical sedimentation.  I describe various mechanisms for particle concentration in gas disks -- including turbulent pressure maxima, drag instabilities and long-lived anticylonic vortices.  I derive a general result for the minimum size for a vortex to trap particles in a sub-Keplerian disk.  Recent numerical simulations  demonstrate that particle clumping in turbulent protoplanetary disks can trigger gravitational collapse.  I discuss several outstanding issues in the field.

\end{abstract}

\maketitle
\section{Introduction}
This chapter serves as a pedagogical introduction to planetesimal formation and related processes in protoplanetary disks.  I defer to \citet{dbcw07} for a comprehensive review of research in this rapidly progressing field.  
Section \ref{s:times} serves as an extended introduction by discussing timescale constraints from disk observations, meteoritics, and dynamics.  The purposeful overlap with other chapters in this volume  aims to reinforce the overlap between sub-fields.  Section \ref{s:mech} describes formation mechanisms: collisional coagulation (\S\ref{s:coag}) and gravitational collapse (\S\ref{s:grav}).  Section \ref{s:drag} covers the dynamics of gas drag in protoplanetary disks, including the basic form of drag forces (\S\ref{s:tstop}), vertical settling and radial migration (\S\ref{s:drsett}), and particle trapping mechanisms (\S\ref{s:trap}).

\section{Timescale Constraints}\label{s:times}
The agreement between T Tauri disk lifetimes and the inferred formation time of meteorites is a remarkable result.  Planetesimal formation appears to be a slow or ongoing process in the several Myr lifetimes of protoplanetary disks.  By contrast the radial migration times of cm to m-sized solids can be under a hundred years!
  
\subsection{Disk Lifetimes}\label{s:disklife}
The standard answer to most questions about disk lifetimes is ``a few million years."  The precise question matters, since main sequence stars (several Gyr old) host low luminosity debris disks (see chapters by Beichman and Kalas).  The Spitzer Space Telescope detects disks with dust masses as low as $\sim 10^{-5} M_\oplus$\citep{mey08}.   Erosive collisions between already formed planetesimals supply the observed debris.

We are concerned here with the lifetimes of optically thick disks from which planetesimals formed.   Near-IR excess emission (i.e.\ above the stellar photosphere) from optically revealed young stellar objects (YSOs) indicates a hot inner disk, and provides the best statistical determination of protoplanetary disk lifetimes.   
 Disk lifetimes are measured by plotting, for many nearby young stellar clusters and moving groups, the fraction of stars with IR excess in a cluster against the average stellar age for that cluster (estimated by comparing the HR diagram to pre-main sequence models).  The disk fraction declines uniformly with age from near unity to near zero, with a (roughly linear) decay time of 5 or 10 Myr (for the H-K or cooler K-L colors, respectively).   See \citet{hil05} for details, caveats, and references.  The fact that we observe a fraction of roughly coeval stars with disks indicates that the scatter among disk lifetimes is also several Myr.  

Millimeter and sub-mm emission traces dust in optically thin outer disk regions, which provides an estimate of disk mass.   The median dust mass in primordial disks (around class I and II YSOs which by definition emit excess IR) is $\langle M_{\rm dust}\rangle \sim 10^{-4} M_\odot \approx 34M_\oplus$,
with some systems ten times as massive \citep{aw07}.  
For solar composition, the total (gas-dominated) mass extrapolates to $\langle M_{\rm disk}\rangle \sim 10^{-2} M_\odot \approx 10 M_{\rm Jup}$.  This is sufficient to form extrasolar giant planets (see the chapters by Udry \& Eggenberger and Guillot), and is consistent with the standard minimum mass solar nebula (MMSN) model, which is itself extrapolated from the solid material in our Solar System \citep{stu77, hay81}.

\citet{aw07} also found that class III YSOs, which lack IR excess, also show little sub-mm emission.  Their average dust mass (dominated by upper limits) is only $\langle M_{\rm dust}\rangle < 5 \times 10^{-7} M_\odot \approx 0.2 M_\oplus$.  This supports using near-IR excess to measure protoplanetary disk lifetimes.

The gas content of disks cannot be weighed as accurately as the dust.  Nevertheless there is ample evidence of gas in protoplanetary disks, especially the hot inner regions \citep{naj07}.  For instance IR excess correlates strongly with accretion diagnostics like H$\alpha$ emission.  We infer that planetesimals form in a gas-rich environment so that drag forces must be included in planetesimal formation theories.  However, there are no strong constraints on dust-to-gas ratios, which need not match solar (or extrasolar host star) compositions.


We conclude that planetesimals should form during the first few Myr of pre-main sequence evolution, when disks have sufficient dust to form terrestrial planets and cores (see the chapter by Ida).  However, the radiative inefficiency of $\gtrsim$ cm-sized solids makes it difficult to observe when planetesimals actually form.

\subsection{Meteoritics}
Solar System (hereafter SS) data provides precise details, unmatched by astronomical observations, but for a single (admittedly special) system.  
Primitive meteorites -- also called chondrites -- preserve early SS history.   Unlike planets and large asteroids, these rocks were never differentiated into an iron-rich core and rocky mantle.  They are relatively unaltered since their time of formation.

Most primitive meteorites contain small ($\lesssim 1$ cm) inclusions which were heated to high temperatures ($\sim 2000$ K), thereby removing volatile elements and turning amorphous dust crystalline.  Spectral signatures of crystalline dust in protoplanetary disks (see chapter by Augereau) show that thermal processing  is ubiquitous.

Chondrules are the most abundant type of inclusion, comprising more than 70\% by volume of the most common primitive meteorites, appropriately called ``ordinary chondrites."   CAIs (calcium aluminum inclusions) are rarer, but are even more refractory (i.e. were subject to a more extreme heating event).  The formation mechanism(s), specifically the heating mechanism, for chondrules and CAIs remains a subject of active research.  The leading hypotheses are: (1) heating near near the young Sun with subsequent outward transport (as in the X-wind model of \citealt{ssl96}), (2) passage though shock waves in the gas disk and (3) collisions between planetesimals or protoplanets.  See the review by \citet{con06} and other chapters in this volume (by Al\'eon and Wood) for details.

CAIs are famous for being the oldest objects with precise ages in the SS.  The record holder is $4567.11\pm0.16$ Myr old, as determined by $^{207}$Pb-$^{206}$Pb dating \citep[][and references therein, hereafter R06]{russ06}.
It is said that CAIs give the ``age of the Solar System,"  whereas they actually date a specific event (the solidification of CAIs) in the history of the Solar System.  There are certainly older events, such as the collapse of the Sun's parent molecular cloud into a protostar and disk \citep{tsc84}, which cannot be dated precisely.  There may even be older rocks; some iron meteorites (which come from the cores of differentiated asteroids) could predate CAIs \citep{tk05}.
The age difference between chondrules and CAIs  sets a lower limit to the formation timescale for primitive meteorites, since the inclusions obviously predate their incorporation into the ``parent body."\footnote{We hope that chondrites are among the first generation of planetesimals formed.  This is not the case if the collisional hypothesis for chondrule formation holds.}

The decay of radioactive $^{26}$Al ($t_{1/2} = 0.73$ Myr) provides evidence that CAIs are older than most chondrules by a few Myr.  Essentially all $^{26}$Al in early solar system rocks has now decayed into stable $^{26}$Mg.  (Only $\sim 10^{-1880}$ of the original $^{26}$Al remains after $4.56$ Gyr!)  Thus short-lived radionuclides (SLRs) like $^{26}$Al are termed ``extinct".   (The chapter by Williams discusses how SLRs, specifically $^{60}$ Fe, $t_{1/2} = 1.5$ Myr, were introduced to the SS.)  The excess of the decay product $^{26}$Mg -- measured by comparing mineral phases with different Al/Mg ratios -- determines the amount of $^{26}$Al in the rock ``initially," i.e. when it last solidified.  Comparing the initial $(^{26}$Al/$^{27}$Al$)_{\rm o}$ ratios 
  between two samples, say a CAI and a chondrule, gives the age difference, $\Delta t$, according to exponential decay:
\begin{equation} \label{eq:dt26Al}
\Delta t = t_{1/2} \ln_2\left[{(^{26}{\rm Al}/^{27}{\rm Al})_{\rm o,1} \over (^{26}{\rm Al}/^{27}{\rm Al})_{\rm o,2}} \right]
\end{equation} 
This technique assumes that $^{26}$Al was uniformly distributed in the disk, otherwise differences in $(^{26}$Al/$^{27}$Al$)_{\rm o}$ could indicate different local formation environment.  For instance the X-wind model generates SLRs near the protostar by spallation reactions, not uniformly in the disk \citep{gou01}.

Most unaltered CAIs share the canonical value $(^{26}{\rm Al}/^{27}{\rm Al})_{\rm o} \approx 4-5 \times 10^{-5}$ (R06).    Applying equation (\ref{eq:dt26Al}) gives a narrow time span, $\sim 0.25$ Myr, for most CAI formation (if the above caveat on uniform mixing is set aside).  By contrast, chondrules with well measured $^{26}$Al (there are only a few) show a range of $(^{26}{\rm Al}/^{27}{\rm Al})_{\rm o} \approx 0.24 \pm 0.17 \times 10^{-5}$ --- $1.4 \pm 0.3 \times 10^{-5}$ (R06).  Equation (\ref{eq:dt26Al}) implies chondrules formed over an extended period of $\sim$ 2 Myr starting $\sim 1$ Myr after CAI formation.

Long-lived radionuclides give absolute ages of early SS rocks by using radioactive isotopes which are not extinct.  The most precise ages come from the $^{207}$Pb-$^{206}$Pb technique, which makes use of the decay chains of $^{235}$U $\rightarrow ^{207}$Pb ($t_{1/2} = 0.704$ Gyr) and $^{238}$U $\rightarrow ^{206}$Pb ($t_{1/2} = 4.47$ Gyr).  
Recent measurements give $^{207}$Pb-$^{206}$Pb ages of $4565.7 \pm 0.4$ Myr, $4564.7 \pm 0.6$ Myr,  and $4562.7 \pm 0.5$ Myr for chondrules in CV, CR and CB meteorite classes, respectively (R06).  These best-fit ages are respectively 1.4, 2.4 and 4.4 Myr after the oldest CAI mentioned above.\footnote{That CAI is from  a CV meteorite, so it's not too surprising that it is closest in age to the CV chondrule.  Note that CV CAIs are large, and thus easier to isotopically date.}

Thus absolute $^{207}$Pb-$^{206}$Pb dates confirm evidence from relative $^{26}$Al chronometry that chondrule formation spanned $\sim 3-5$ Myr after the formation of CAIs.   R06 summarize the evidence in more detail, including  cautionary notes.  Thus planetesimal formation was either delayed or ongoing for several Myr.  This age spread fits within the observed lifetimes of protoplanetary disks.  Rapid planetesimal formation in $< 1$ Myr is not supported by the data.

\subsection{Radial Migration by Aerodynamic Drag}\label{s:driftproblem}
Disk dynamics also constrains the time it takes for planets to form.  Planets, and their building blocks, have the inconvenient tendency to fall into the star.   The migration of planets is a significant issue for their survival and orbital evolution, especially for hot Jupiters.  However the aerodynamic migration of small solids near the meter-size barrier
imposes the most stringent timescale constraint on planet formation: $\approx 100$ years.  (This is rivaled only by thesis, job and grant deadlines!) Moreover the  aerodynamic drift mechanism is simple and robust, requiring only that the orbital motion of the gas disk is slightly sub-Keplerian due to gas pressure.  Disk models are generically hotter and denser closer to the YSO, so the requirement of a globally decreasing pressure gradient is almost certainly met.  

A solid particle meets a headwind as it orbits in a sub-Keplerian gas disk.\footnote{If the particle orbit is sufficiently eccentric ($e > 2 \eta - \eta^2$ to be precise) then it will experience a tailwind near perihelion.  However eccentricities (and inclinations) not only damp more rapidly than semi-major axes, but also increase the orbit-averaged rate of in-spiral \citep{ahn76}.}  This headwind saps the particle of angular momentum causing it to inspiral towards the star. 
Local pressure perturbations (including vortices) can temporarily halt the inflow in special locations  (see \S\ref{s:trap}). However a net influx will continue with a reduced rate (an order unity correction in \citealp{jkh06}) that depends on the efficiency and duration of the particle ``traps."

 The amplitude of the sub-Keplerian gas rotation, $\delta v_{\phi, {\rm g}} \equiv v_{\phi, {\rm g}} - v_{\rm K}$ (with $v_{\rm K} \equiv \sqrt{GM_\ast/R} = \varOmega R$ the Keplerian orbital speed at the cylindrical radius, $R$), is readily found by balancing the force from the radial pressure gradient, $f_{P,R} = -\rho_{\rm g}^{-1}\partial P/\partial R$ (where $\rho_{\rm g}$ and $P$ are the gas density and pressure), with the Coriolis force, $f_{{\rm Cor},R} = 2\varOmega \delta v_{\phi, {\rm g}}$.  Setting $f_{P,R} + f_{{\rm Cor},R} = 0$ gives:
\begin{equation}\label{eq:dvphig}
\delta v_{\phi, {\rm g}} = {1 \over 2 \rho_{\rm g} \varOmega} {\partial P \over \partial R} 
\equiv -s_P  {c_{\rm s}^2 \over v_{\rm K}} \equiv -\eta v_K \approx - 56 \, {\rm m \over s} \, ,
\end{equation} 
where $c_{\rm s} \equiv \sqrt{P/\rho_{\rm g}}$ is the isothermal sound speed, and $\eta$ is the dimensionless measure of pressure support. 
The factor $s_P  \equiv - (\partial  \ln P/\partial \ln R)/2$ depends only on the slope of the gas pressure, with $s_P \approx 1.6$ in Hayashi's MMSN model.  Thus $\delta v_{\phi, {\rm g}}$ is independent of disk mass and depends mainly on the midplane temperature, $T \propto c_{\rm s}^2$.    Pressure support is weaker, $\delta v_{\phi, {\rm g}} \approx -30 (R/{\rm AU})^{1/14}$ m/s, for passively-irradiated flared-disk models \citep[][hereafter CG97]{cg97}\footnote{Since CG97 derive $T \approx 150 (R/{\rm AU})^{-3/7}$ K vs.\ $T = 280 (R/{\rm AU})^{-1/2}$ K for the MMSN.}, but stronger in active accretion disks.

The peak radial drift speed  of a particle is simply $\delta v_{\phi, {\rm g}}$, directed inwards, for particles whose aerodynamic coupling time equals the orbital time (see \S\ref{s:drsett}).  Meter-diameter compact solids have the fastest drift speeds at 1 AU in the MMSN, while smaller solids drift fastest in the outer disk, e.g.\  3 cm at 35 AU.  While the \emph{size} of the fastest migrating particle varies with gas density, the peak drift \emph{speed} does not.   Aerodynamic drag can bring a particle toward the star in a drift time
\begin{equation} 
t_{\rm drift} \equiv R/\max (v_{\rm drift}) \approx 85 \,(R/{\rm AU})~ {\rm yr}
\end{equation} 
for the MMSN model, or $t_{\rm drift} \approx 160~(R/{\rm AU})^{13/14}$ yr for the CG97 model.\footnote{The drift or migration timescale, $t_{\rm mig} = R/|v_R|$ (with $v_R$ the local radial drift speed) is the characteristic time to travel of order the current distance to the star.   The time to actually reach the star -- the integral quantity $\int_{R_o}^{R_\ast}dR/v_R(R)$ --  depends on the form of $v_R(R)$.}  Avoiding catastrophic loss of planetesimal forming material requires rapid growth though the fastest drifting sizes -- direct collapse could skip the problematic size range entirely -- and/or efficient trapping of particles (\S\ref{s:trap}).

For comparison, I briefly review planetary migration (see also chapters by Terquem and Ida). ``Type I" migration occurs because a planet launches spiral waves in the gas disk.  The gravitational back-reaction of the waves on the planet exerts tidal torques, which result in a net inward migration \citep{war97}.  The migration timescale from the 3D calculation by  \citet{ttw02} is:
\begin{equation}\label{eq:migI}
t_{\rm mig, I} \approx 0.22 {M_\ast \over M_{\rm pl}}{M_\ast \over \varSigma_{\rm g} R^2} \left({H_{\rm g} \over R}\right)^2 \varOmega^{-1} \approx 3 \times 10 ^5 {\rm yr}\left({M_{\rm pl} \over M_\oplus}\right)  \left({R \over 5\,{\rm AU}}\right)^{3/2}\, ,
\end{equation} 
where $H_{\rm g} = c_{\rm s}/\varOmega$ is the gas scaleheight.  Migration is faster if either the planetary mass,  $M_{\rm pl}$, or the characteristic  disk mass, $\varSigma_{\rm g} R^2$ (the MMSN is assumed in the estimate), increases relative to the stellar mass $M_\ast$.  However, there is a limit. Tidal torques can overwhelm disk viscosity and open an annular gap in the disk when the planet mass exceeds \citep{lp86}:
\begin{equation} \label{eq:mgap}
M_{\rm pl} > \sqrt{40 \alpha}\left({H_{\rm g} \over R} \right)^{5/2} M_\ast \approx 0.3 M_{\rm Jup} \sqrt{\alpha \over 10^{-2}} \left({R \over 5\,{\rm AU}}\right)^{5/8} \, ,
\end{equation} 
 where $\alpha$ is the dimensionless parameterization of the disk's turbulent viscosity $\nu \equiv \alpha c_{\rm s}^2/\varOmega$.  Applying inequality (\ref{eq:mgap}) to equation (\ref{eq:migI}) gives an upper limit to the type-I migration rate of
 \begin{equation} 
t_{\rm mig, I} <  0.22 {M_\ast \over \varSigma_{\rm g} R^2} \sqrt{R \over 40 \alpha H_{\rm g}  } \approx 7000 ~{\rm yr}\sqrt{10^{-2} \over \alpha} \left({R \over 5\,{\rm AU}}\right)^{7/8}
\end{equation} 
Once a planet opens a gap, ``type II" migration occurs on the viscous  timescale,
\begin{equation} 
t_{\rm mig, II} \approx {R^2 \over \nu} = {1 \over \alpha \varOmega} \left(R \over H_{\rm g} \right)^2 \approx  7 \times 10^4~{\rm yr} \, {10^{-2} \over \alpha}  \left({R \over 5\,{\rm AU}}\right)\, .
\end{equation}
and is generally directed inward (following the flow of the accretion disk). 
 
 This summary glosses over the many uncertainties and difficult technical issues associated with planet migration.  For instance the horseshoe orbits of gas in the corotational region of the planet could lead to a rapid type III migration whose direction depends on initial conditions \citep{pam08}.  We conclude that aerodynamic drift of small solids (sometimes called type-0 migration) is both more rapid and based on better understood processes than the migration of planets.

\section{Growth Mechanisms}\label{s:mech}
We now turn to the proposed mechanisms for the formation of planetesimals: (1) successive collisions between solids which result in agglomeration and (2) the gravitational collapse of smaller solids into a larger planetesimal.  The two mechanisms are not mutually exclusive: dust grains might grow by sticky collisions until solids decouple sufficiently from the gas to collapse under their self-gravity.

\subsection{Collisional Coagulation}\label{s:coag}
Collisions between particles can lead to the rapid growth of planetesimals (including comets in the outer solar system) if particle sticking is efficient \citep{stu80, stu95}.  The key (and difficult) issue is understanding when collisions yield particle growth, as a function of particle sizes, impact speeds, orientation and material properties.  
Significant progress is being made in the laboratory, including microgravity experiments in drop towers, sounding rockets and the space shuttle \citep[see][hereafter BW08 for a comprehensive review]{bw08}.

Relative velocities between particles regulate collision rates and sticking probabilities.  Turbulent stirring, aerodynamic drift (both radial and azimuthal) and vertical settling contribute to collision speeds during planetesimal formation.  The speeds increase monotonically with particle size, until decoupling occurs at $\sim$ meter-sizes, when relative speeds reach $\sim$ 100 m/s.  Fig.\ 4 of \citet[see references therein for details]{dbcw07} shows this clearly.

Collisional growth requires (1) a binding energy (BE) which is significant compared to the kinetic energy (KE) and (2) a sufficient fraction of  KE dissipation, $f_{\rm diss}$, during the impact.  Objects should stick when the initial KE obeys
\begin{equation} \label{eq:KEBE}
{\rm KE_{-\infty}} < |{\rm BE}|  {f_{\rm diss} \over 1 - f_{\rm diss}}\, .
\end{equation}
The logic, illustrated in Fig.\ {\ref{f:impact}, is as follows.  At the moment of impact, the KE increases to ${\rm KE_{-\infty}} +  |{\rm BE}|$ by falling into the potential well  (it's easy to visualize smooth acceleration in a gravitational potential, but contact forces work  also).  Dissipation during the impact then gives ${\rm KE} \rightarrow (1-f_{\rm diss})({\rm KE_{-\infty}} + |{\rm BE}|)$.  Climbing back out of the potential well gives a final kinetic energy ${\rm KE_{+\infty}} = (1-f_{\rm diss})({\rm KE_{-\infty}} + |{\rm BE}|) - |{\rm BE}|$.  Imposing the condition to remain bound, ${\rm KE_{+\infty}} < 0$, reproduces equation (\ref{eq:KEBE}).  This heuristic model formalizes the idea that sticking requires some combination of gentle collisions, strong binding and energy dissipation.  Collisions between a small projectile (which dominates the center-of-mass kinetic energy) and a large target (which contributes to the potential energy and provides a sink for dissipated energy) are especially favorable for growth.
A cruder (but useful) approximation is that coagulation occurs when ${\rm KE_i} \lesssim |{\rm BE}|$, which assumes that $f_{\rm diss}$ is not very close to either 0 or 1.

\begin{figure}
\vspace{-0.5cm}
\begin{center}
\includegraphics[width=12cm]{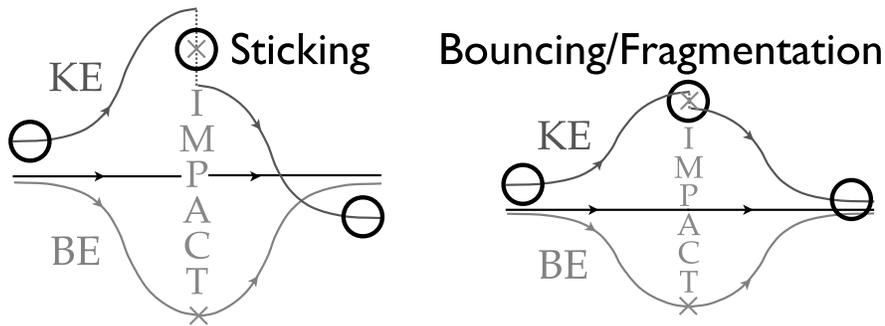}
\end{center}
\vspace{-0.5cm}
\caption{A schematic illustration showing the evolution of kinetic (KE) and potential or binding energy (BE) during a collision.  Sticking  requires dissipation of KE so that objects remain bound.  Escape to infinity in the \emph{left} example would require negative KE, and thus does not occur.  Bouncing or fragmentation (\emph{right}) occurs if there is less dissipation and objects escape with positive KE.  See equation (\ref{eq:KEBE}) and surrounding  text.}\label{f:impact}
\end{figure}

The available binding energies are electrostatic and gravitational interactions. Short-range van der Waals interactions explain the sticking of individual grains and small aggregates at speeds $\lesssim 1$ m/s \citep[][]{dt97}. 
   As particles increase from $\mu$m to m-sizes, coagulation is more difficult since: (1) the available BE declines as the surface area-to-volume ratio drops  \citep[][]{you04}, while simultaneously (2) collisional KE rises as particles decouple from the gas.   Net (and opposite) charges can help, but feasibility has not been demonstrated.  Experiments find that mm-sized aggregates do not stick at any speed, with bouncing (fragmentation) for impacts above (below) $\sim$ m/s (BW08).    

Gravitational BE is only significant once massive planetesimals have formed.  The ${\rm KE_i} \lesssim |{\rm BE}|$ criterion is equivalent to requiring (for roughly equal masses) that the collision speed is less than the surface escape speed: $v_{\rm coll} \lesssim v_{\rm esc} \sim 1\, (a/{\rm m}) ~{\rm mm/s}$, here normalized to a compact particle with radius $a = 1$ m.  This velocity scale is orders of magnitude too small to account for sticking since $v_{\rm coll} \gtrsim 10$ m/s due to aerodynamic drift alone  (assuming a modest dispersion in sizes).

So what about the scenario of small projectiles sticking to a few ``runaway" bodies that become planetesimals?  The population of small bodies could be maintained by erosive collisions between large bodies \citep{jbd08}.  However collisions of small bodies ($\ll$ m-sized and tied to the sub-Keplerian gas) with large solids (on decoupled Keplerian orbits) face the full brunt of sytematic azimuthal drift, $\delta v_{\phi, {\rm g}} \gtrsim 50$ m/s, plus a fluctuating contribution from stirring by turbulent gas.  
Since these speeds are comparable to a sand-blaster (credit to Eugene Chiang for this analogy)  it is difficult to imagine net growth.  BW08 conclude conservatively that; ``The direct formation of kilometer-sized planetesimals cannot (yet?) be understood via sticking collisions."  This author concurs.

Many collision models predict the growth of bodies $\gtrsim$ cm in $\sim 1000$ yr (BW08 and references therein).   \citet{dd05} make the important point that such rapid growth leads to a premature decline in excess IR and mm emission, which would contradict observed Myr disk lifetimes (\S\ref{s:disklife}).  Fragmentation (and/or inefficient sticking in the first place) is needed to maintain a population of small grains.  Astronomical observations are a powerful constraint on our imperfect knowledge of collisional physics.

\subsection{Gravitational Collapse}\label{s:grav}
If the ``bottom up" growth of planetesimals by collisional coagulation stalls, then a ``top down" phase  is needed, during which self-gravity fragments the disk of small solids into planetesimals.    Gravitational instability (GI)  is responsible for the formation of most objects in the universe, including cosmological structure and stars.  It has been hypothesisized that giant planets can form by gravitational fragmentation of protoplanetary gas disks \citep{boss97}.  However the gas is unlikely to cool sufficiently rapidly to collapse into planets \citep{raf07}.  Here we are concerned with the collapse of the \emph{solids} in protoplanetary disks, which cool (i.e. damp their random kinetic energy) rapidly by gas drag and inelastic collisions.

\subsubsection{Stability Criteria}

\citet{saf69} and \citet[][hereafter GW73]{gw73} independently proposed that planetesimals form by GI.  They applied two standard criteria for gravitational collapse: the Roche (or critical) density and the \citet{too64} $Q$-parameter.  The Roche limit is usually invoked for tidal disruption, e.g. the rings inside Saturn's Roche lobe, but collapse is merely the inverse of disruption.  Gravitational collapse is possible when the self-gravity, with an acceleration $f_{\rm sg} \sim G\rho \ell$ for a region of density $\rho$ and size $\ell$, exceeds the differential tidal gravity, $f_{\rm tide} \sim GM_\ast \ell /R^3$, at a distance $R$ from the central mass, $M_\ast$.  This gives the simple collapse criterion $\rho > \rho_{\rm R} \sim M_\ast/R^3$, but no information about the characteristic lengthscale since we ignored pressure stabilization of small scales.

The Toomre criterion states that thin disks are gravitationally unstable when
\begin{equation} \label{eq:QT}
Q_{\rm T} \equiv {c \varOmega \over \pi G \varSigma}  <  1
\end{equation} 
where $c$ is the random velocity (for a gas the sound speed), $\varOmega$ is the orbital frequency\footnote{More generally the epicyclic frequency, $\kappa$, should appear, but $\kappa = \varOmega$ in a Keplerian disk.}, and $\varSigma$ is the surface density (see \citealp{shuv2} or \citealp{bt08} for a derivation).  This criterion says that mass promotes gravitational collapse, while pressure and epicyclic restoring forces (represented by $c$ and $\varOmega$ respectively) stabilize the disk.  The dispersion relation (see above texts) restricts the unstable wavelengths to be less than $\lambda_{\rm T} = 4\pi^2 G \varSigma/\varOmega^2$, since angular momentum conservation prevents large regions from collapsing.  The Roche criteria can be rearranged to look like the Toomre criteria:
\begin{equation} \label{eq:QR}
\rho \gtrsim {M_\ast \over R^3} \Leftrightarrow Q_{\rm R} \equiv {H \varOmega^2 \over \pi G \varSigma} \lesssim 1
\end{equation} 
where the scaleheight, $H \approx \varSigma / \rho$.  Equations (\ref{eq:QT}) and (\ref{eq:QR}) are roughly equivalent for $c \approx \varOmega H$, which holds for gas disks and most particle orbits, but does not hold for small particles suspended in the gas disk \citep[e.g.][]{yl07}.
 
The standard GI criteria (eqs.\ [\ref{eq:QT}] and [\ref{eq:QR}]) apply to  \emph{single component} disks. In the presence of gas drag,  collapse is possible even when $Q_{\rm T}$ and $Q_{\rm R} \gg 1$  \citep[though the basic idea is in GW73]{war00},\footnote{These models do include radial turbulent diffusion, which slows but does not prevent this mode of collapse (Youdin, in prep).} The transfer of (excess) angular momentum, from the particle layer to the  gas, allows longer wavelengths to collapse.  Since self-gravity is weaker, these dissipative gravitational instabilities can be slow, especially for small particles that must push through the gas.

\subsubsection{Turbulence: A Curse and a Cure}\label{s:turbGI}
The stirring of small solids by turbulent gas is the fundamental obstacle to gravitational collapse.  In the absence of any turbulence, particles would settle to arbitrarily high densities in the midplane -- faster than they drift radially -- so that GI is trivial.
Hoping that turbulence disappears, even just in the midplane, is not a very satisfactory solution.  For instance, turbulent stresses can propagate into ``dead zones"  from the active surface layers of accretion disks \citep{omm07}.

Moreover, particle settling  toward the midplane robustly generates midplane turbulence.  As solids sediment into a dense layer, they drag the midplane gas along at faster (closer to Keplerian) speeds, generating vertical shear with the overlying sub-Keplerian gas.  This shear in turn triggers Kelvin-Helmholz instabilities (KHI) which develop into turbulence.  The presence of KHI was recognized by GW73 and noted to be an obstacle to GI by \citet{stu80}.  Research on KHI in protoplanetary disks continues \citep[e.g.\ ][]{chi08}. 

Particles can also generate midplane turbulence via streaming instabilities, hereafter SI \citep{yg05}.   Streaming motions between solids and gas --  the naturally occuring radial and azimuthal drift  -- supply the free energy for  SI.  By contrast gas velocity gradients, i.e.\ shear, are responsible for KHI.  To model SI, the gas and solid components must evolve independently, either as two fluids, or with a hybrid hydrodynamics and N-body scheme \citep{yj07}.
A hybrid of KHI and SI (simulated by \citealp{jhk06} in 2D and \citealp{nature07}, hereafter J07, in 3D) operates in real disks with stratification and imperfect particle-gas coupling (i.e.\ with slippage allowed).

Midplane turbulence may not be the fatal flaw for GI, for two reasons.  Both effects are augmented in disks with high metallicity (and thus large initial solid abundances), consistent with the observation that exoplanets are more common around metal-rich stars \citep{fv05}.
First, KHI can only stir finite amounts of solids \citep{sek98,ys02}.  If the local abundance of solids relative to gas exceeds $\sim 10$ times solar abundances, then any excess solids cannot be stirred by vertical shear, and will be subject to GI (absent sufficient stirring by other sources of turbulence).  This saturation limit exists because the amount of destabilizing vertical shear is limited (by $\delta v_{\phi, g} \approx 50$ m/s) while the stabilizing anti-buoyancy increases as more solids are poured into the midplane.

Second, while turbulence diffuses particles in an average sense, it can also generate intermittent particle clumps.  Enhanced self-gravity in an overdense clump is conducive to GI.  Mechanisms for particle concentration in turbulence (broadly defined to include longer-lived vortices) are described in \S\ref{s:trap}.  Particle clumping is particularly strong for SI \citep{jy07}, since particle overdensities generate perturbations to the drag force that sustain the growth of SI.

The models of J07 demonstrate the viability of planetesimal formation by GI in turbulent disks.   Their hybrid MHD/particle-mesh simulations include several sources of turbulence -- vertical stratification (for KHI), particle-gas streaming (for SI), and magnetic fields to drive the magneto-rotational instability (MRI).\footnote{The supplement to J07 shows that GI occurs in the absence of magnetic fields, though higher particle surface densities were needed.  This is surprising,  since (naively) less turbulence should favor GI.  The clumping in the MRI turbulence strengthens the feedback of drag forces on the gas, which in turn  enhances the concentration by SI.}  Particle clumps grew in the absence of self-gravity, and then collapsed rapidly when self-gravity was included.  The largest bound clump accreted at an astonishing rate of $\sim 0.5 M_{Ceres}$ per orbital period.  A clump mass that exceeds a nominal km-sized planetesimal by orders of magnitude is not a serious concern, since hierarchical fragmentation in the later stages of collapse (not yet modeled) is plausible.

The initial particle sizes in J07 are in the 15---60 cm range, making them moderately decoupled from the gas.   Growth by coagulation to these sizes is not certain (\S\ref{s:coag}).  Strengthening the GI (or Safronov-Goldreich-Ward) hypothesis requires (1) a better understanding of particle growth beyond $\sim 10$ cm and/or (2) pushing the dynamical models to smaller initial particle sizes, which should give slower growth rates and may require higher initial particle abundances.  

\begin{figure}
\vspace{-0.2cm}
\hspace{2cm}
\includegraphics[height=6.5cm]{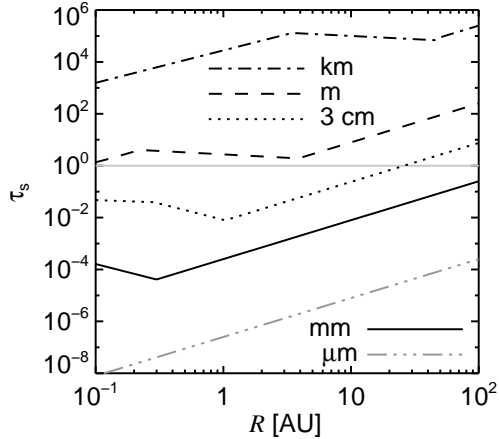}
\vspace{-0.5cm}
\caption{Dimensionless stopping time, $\taus$, vs.\ disk radius in the midplane of a minimum mass nebula model for particle sizes from $\mu$m -- km. The sharp transitions between different drag regimes are approximations.}\label{f:ts}
\end{figure}

\section{Aerodynamic Coupling in Particle-Gas Disks}\label{s:drag}
\subsection{Drag Laws and the Stopping Time}\label{s:tstop}
While planets interact with gas disks  gravitationally, smaller solid bodies are coupled to the gas disk primarily by drag forces.  The drag force, ${\bf F}_{\rm d}$, on a particle of mass $m_{\rm p}$ moving with a speed $\Delta \vc{v} \equiv \vc{v}_{\rm p} -  \vc{v}_{\rm g}$ relative to the gas is
\begin{equation}\label{eq:Fdrag}
{\bf F}_{\rm d} = -m_{\rm p} {\Delta \vc{v} / \ts}\, ,
\end{equation} 
where $\ts$ is the damping timescale for particle motion relative to the gas.  The form of $\ts$ depends on particle  properties  -- such as the internal density, $\rho_{\rm s}$, and spherical radius, $a$ -- and on properties of the gas disk --  $\rho_{\rm g}$ and $c_{\rm s}$ -- as:
\begin{subnumcases}{\ts = }
\ts^{\rm Ep} \equiv \rho_{\rm s} a / (\rho_{\rm g} c_{\rm s}) & if  $a < {9 \lambda/4}$\label{Ep} \\
\ts^{\rm Stokes} \equiv \ts^{\rm Ep} \cdot  4a/(9 \lambda) & if  ${9 \lambda/4} < a < \lambda /(4 {\rm Ma})$ \label{St}\\
\ts^{\rm Ep}\cdot \left(a / \lambda\right)^{3/5} {\rm Ma}^{-2/5}/4 & if $\lambda /(4 {\rm Ma}) <a<200 \lambda/{\rm Ma}$ \label{mix} \\
\ts^{\rm turb} \equiv \ts^{\rm Ep} \cdot 6 /{\rm Ma} & if  $a > 200 \lambda/{\rm Ma}$ \label{turbdrag}
\end{subnumcases}
where ${\rm Ma} \equiv |\Delta\vc{v}|/c_{\rm s}$, $\lambda \propto 1/\rho_{\rm g}$ is the gas mean free path, and  ${\rm Re} \equiv 4 a {\rm Ma}/\lambda$ will be the Reynolds number of the flow around the particle.  The cases are written in order of increasing particle size: Epstein's Law of drag from molecular collisions, Stokes' Law for viscous drag when ${\rm Re}  < 1$, an approximate intermediate ${\rm Re}$ case, and the drag from a fully developed turbulent wake for ${\rm Re} > 800$.

We measure the dynamical significance of drag forces with the parameter,
\begin{equation} 
\taus \equiv \Omega \ts\, .
\end{equation} 
For $\taus \ll 1$ particles are carried along with the gas, while for $\taus \gg 1$ gas drag is a small correction to Keplerian orbits.
Fig.\ \ref{f:ts} plots $\taus$ values for several particle sizes in the midplane of the MMSN with a gas surface density, $\varSigma_{\rm g} = 2000 (R/{\rm AU})^{-3/2}$ ${\rm g/cm}^2$, and midplane temperature, $T = 280  (R/{\rm AU})^{-1/2}~{\rm K}$.
The relative velocities needed for the turbulent drag formulae (eqs.\ [\ref{mix},\ref{turbdrag}]) are determined self-consistently by inplane drift (eq.\ [\ref{eq:Dv}]).  Turbulent stirring and vertical oscillations will give a small correction.

\begin{figure}
\vspace{-0.5cm}
\hspace{-0.5cm}
\includegraphics[height=6.5cm]{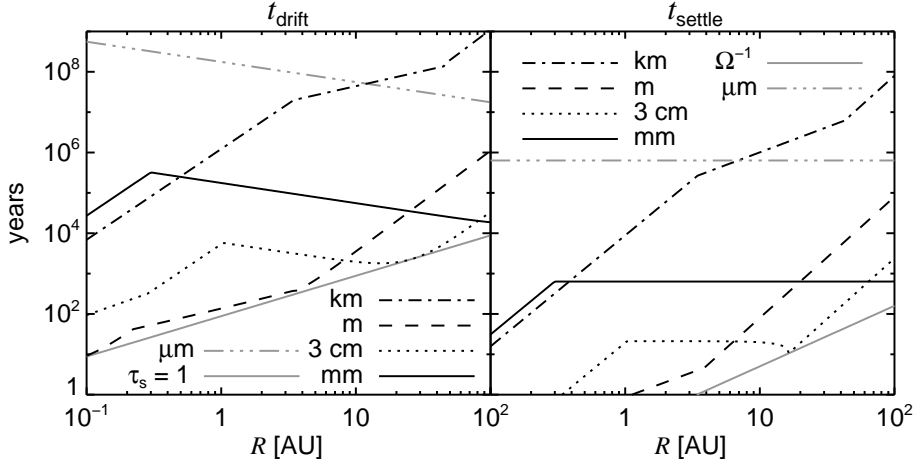}
\vspace{-0.9cm}
\caption{ (\emph{Left:}) Particle radial drift timescale for several sizes vs.\ disk radius.  The fastest migration at any radius is for a particle with $\taus = 1$ (\emph{grey curve}).  (\emph{Right:}) Particle settling times, which are longer than the orbital time (\emph{grey curve}), but shorter than the radial drift times.}\label{f:tdrsett}
\end{figure}

\subsection{Settling and Drift}\label{s:drsett}
\subsubsection{In-plane Drift}\label{s:drift}
We now derive the particle drift motions.  First the orbital speed of the gas, $v_{\phi, {\rm g}}$, is given by centrifugal balance with the Keplerian gravity and pressure forces,
\begin{equation} 
-{v_{\phi, {\rm g}}^2 / R} = -{v_{\rm K}^2 / R} - \rho_{\rm g}^{-1} {\partial P / \partial R}\, .
\end{equation}  
We  ignore the back reaction of drag forces on the gas, valid in the test-particle limit when the density of particles $\rho_{\rm p} \ll \rho_{\rm g}$. 
When pressure forces are small compared to radial gravity\footnote{Otherwise the disk would be spherical!}  the approximate solution,  $\delta v_{\phi, {\rm g}} \approx v_{\phi,{\rm g}} - v_{\rm K} =  - \eta v_{\rm K} \approx - 56 ~{\rm m/s}$, matches equation (\ref{eq:dvphig}).
Rotation is again sub-Keplerian for the (usual) outwardly decreasing pressure, with $\eta \equiv - (\partial P/ \partial R)/(2\rho_{\rm g} \varOmega^2 R) \approx (c_{\rm s}/v_{\rm K})^2 \sim 10^{-3}$.

The particle response satisfies the radial and azimuthal equations of motion:
\begin{subequations}\begin{eqnarray}
\ddot{R} - R \dot{\phi}^2 &=& -v_{\rm K}^2 / R - \dot{R} / \ts \\
R \ddot{\phi} + 2 \dot{R} \dot{\phi} &=& -(R\dot{\phi} -v_{\phi,{\rm g}}) / \ts
\end{eqnarray}\end{subequations}
where the radial gas velocity vanishes.  For steady state solutions, we set $\ddot{R} \approx 0$ and let $\dot{\phi} \equiv \varOmega(R) + \delta v_{\phi, {\rm p}}/R$ so that $R\ddot{\phi} \approx -(3/2) \varOmega \dot{R}$.  Using equation (\ref{eq:dvphig}) for the gas velocity gives the particle drift speeds,
\begin{eqnarray} 
v_{R, {\rm p}}  = \dot{R} = -  {2 \taus \eta v_{\rm K} \over 1 + \taus^2}\, 
\end{eqnarray} 
and $\delta v_{\phi,{\rm p}} = - \eta v_{\rm K}/( 1 + \taus^2)$.  The inward drift speed is maximized at $-\eta v_{\rm K}$ for $\taus = 1$ particles.  The total (in-plane) particle speed relative to gas, 
\begin{equation} \label{eq:Dv}
|\Delta\vc{v}| = \sqrt{\dot{R}^2 + \left(\delta v_{\phi,{\rm p}} + \eta v_{\rm K}\right)^2 } = {\sqrt{4\taus^2 + \taus^4} \over 1 + \taus^2} \eta v_{\rm K}
\end{equation}
asymptotes to $ |\Delta\vc{v}| = \eta v_{\rm K}$ for large particles ($\taus \gg 1$) on Keplerian orbits.

Fig.\ \ref{f:tdrsett} (\emph{left panel}) plots the radial drift timescale, $t_{\rm drift} \equiv R/|v_{R, {\rm p}}|$ for a range of particle sizes.  The fastest drift,  $t_{\rm drift, min} = (\eta \Omega)^{-1} = 88 (R/{\rm AU})$ yr is for $\taus = 1$ particles (\emph{shown in grey}).

\subsubsection{Vertical Settling}
Vertical particle motion decouples from in-plane motion, obeying the simple ODE,
\begin{equation} 
\ddot{z} = -({\dot{z} - v_{z,{\rm g}} )/ t_{\rm stop}} - \varOmega^2 z\, ,
\end{equation} 
where $g_z \approx - \varOmega^2 z$ is the Keplerian vertical gravity near the midplane ($z \ll R$).  When forcing by the vertical gas velocity, $v_{z,{\rm g}}$, vanishes (for a laminar disk) we have a damped harmonic oscillator, with a damping -- i.e.\ particle settling -- time:
\begin{equation} 
t_{\rm sett} = {2 \ts \over \rm{Re} [1 - \sqrt{1-4 \taus^2}]} \approx {1 + 2 \taus^2 \over \varOmega \taus}\, .
\end{equation} 
The final, approximate expression is a simple matching of the tight and loose coupling limits.  For $\taus \ll 1$, overdamped oscillations bring particles directly to the midplane at their terminal velocity $v_{\rm sett}(z) = g_z \ts = -\varOmega^2 z \ts$, giving a settling time $t_{\rm sett} = z/v_{\rm sett}(z) = (\Omega^2 \ts)^{-1}$.  For large particles with $\taus \gg 1$, oscillation amplitudes are damped as $\exp[-t/(2\ts)]$.  Fig.\ \ref{f:tdrsett}  (\emph{right panel}) plots settling times for several particle sizes.  Note that $\mu$m-sized grains have  $t_{\rm sett} \approx$ Myr, of order the gas disk lifetime, emphasizing the need for some particle coagulation.

Forcing by the turbulent gas maintains a finite particle layer thickness, $H_{\rm p}$. This is given by balancing $t_{\rm sett}$ with the turbulent diffusion timescale across $H_{\rm p}$: $t_{\rm diff} \sim H_{\rm p}^2 / D_{\rm p}$.   We skip details  of how the particle diffusion coefficient, $D_{\rm p}$, depends on $\taus$ \citep{yl07}.   The particle scaleheight is given by a single simple expression \citep{cfp06, yl07},
\begin{equation} 
H_{\rm p} \approx \min\left(\sqrt{\alpha_{\rm g}/\taus},1\right) H_{\rm g}\, ,
\end{equation} 
where $\alpha_{\rm g} \equiv D_{\rm g}/(H_{\rm g}^2 \varOmega)$ parametrizes turbulent diffusion in the gas.\footnote{This $\alpha_{\rm g}$ is  not identical -- but can be similar in magnitude --  to the usual  $\alpha$ for angular momentum transport, which includes Maxwell stresses.} The $\sqrt{\alpha_{\rm g}/\taus}$ factor tells us that weaker turbulence and/or larger, heavier particles result in a thinner particle later.  Taking the minimum simply  prevents $H_{\rm p} > H_{\rm g}$.

\subsubsection{Small Particles Move (Relatively) to High Pressure}
A simple rule of thumb is that particles move to regions of high pressure.  Midplane settling and inward radial drift (or falling rain\footnote{Sadly there was little snow during the Winter School.} here on Earth) follow this pattern.   To derive this result (see also \citealp{jkh06}), and get a simple prescription for the drift speeds of small particles, consider the equations of motion for the gas velocity, $\vc{v}_{\rm g}$, and a particle labeled by index $i$:
\begin{subequations}\label{eq:balance}\begin{eqnarray} 
{D \vc{v}_{\rm g} \over D t} -{\bf  f}_{\rm ext}(\vc{x}) &=&  -{\nabla P \over \rho_{\rm g}}\\
\ddot{\vc{x}}_i -{\bf  f}_{\rm ext}(\vc{x}_i) &=& -{\dot{\vc{x}}_i - \vc{v}_{\rm g}(\vc{x}_i) \over \ts} \equiv -{\Delta \vc{v}_i \over \ts}
\end{eqnarray}\end{subequations}
Whenever $\taus \ll 1$, particles are frozen-in to the gas to lowest order, so that $ \dot{\vc{x}}_i \approx \vc{v}_{\rm g}(\vc{x}_i) + \mathcal{O}(\taus)$.  Thus the left hand sides of eq.\ [\ref{eq:balance}a,b] are in balance, including external forces, ${\bf f}_{\rm ext}$ -- such as gravity or Coriolis forces -- which affect both components.  Thus the ``extra" force felt by the gas -- pressure gradients -- must balance the ``extra" force felt by the particles -- drag.\footnote{Pressure (also, magnetic fields, ignored here) has a negligible effect on dust grains and solids because of their large ``molecular weight."   Drag forces have little effect on the gas in the test particle limit when the mass density of particles $\rho_{\rm p} \ll \rho_{\rm g}$.}   Consequently
\begin{equation} \label{eq:driftrule}
\Delta \vc{v}_i \approx {\nabla P \over \rho_{\rm g}} \ts + \mathcal{O}(\taus^2)\, ,
\end{equation} 
which gives the correct radial drift and vertical settling speeds, but not the order $\taus^2$ azimuthal drift speed.   We apply this result to vortex trapping (eq.\ [\ref{eq:divvort}]).

\subsection{Particle Trapping Mechanisms}\label{s:trap}
A promising route to planetesimal formation involves the clumping of particles by gas structures in the disk, which are mostly turbulent in origin.  Particle density enhancements can then seed gravitational collapse (\S\ref{s:grav}).  These trapping mechanisms also slow the migration of solids, a serious problem in its own right (\S\ref{s:driftproblem}).

\subsubsection{Pressure Reversal}
The tendency for particles to drift radially inward is again due to the azimuthal headwind caused by the global radial pressure gradient, $\partial P_o/\partial R \sim - \rho_{\rm g} c_{\rm s}^2 /R < 0$ (the subscript  ``o"  now indicates a smoothing of fluctuations on scales $\ell \ll R$).   A stationary point, with no headwind or radial inflow, exists wherever a local pressure perturbation, $P'$, flattens the total radial pressure gradient,
\begin{equation} 
\left.\partial P/\partial R\right|_{\rm total} = \partial P_o/\partial R + \partial P'/\partial x = 0\, ,
\end{equation} 
with $x$ the local radial coordinate.  Pressure maxima (not minima) are stable stationary points.  Small pressure perturbations, $P' \ll P_o$, reverse the gradient if they vary on length scales, $\ell \lesssim R P'/P_o \ll R$.  A lower limit, $\ell \gtrsim \ell_{\rm stop} = \dot{R}_o \ts = 2 \eta R \taus^2/(1+\taus^2)$, ensures that particles can stop at the stationary point.   Long-lived fluctuations are more effective traps, with $\ts$ the minimum duration.  The fastest drifters (with $\taus = 1$) concentrate most efficiently when these criteria are met.  

What structures can generate the required pressure maxima?
Self-gravitating gaseous spiral arms are one possibility \citep[e.g.\,][]{ric04}.  The pressure perturbations must be substantial for large (AU-scale) density waves (applying the gradient argument above).   This scenario requires a massive, and presumably young disk, which may not be favored according to the arguments in (\S\ref{s:times}).

Pressure fluctuations in disk turbulence are another candidate.  Simple scalings give the pressure perturbations in terms of the turbulent Mach number and $\alpha$ as: $P'/P_o \sim {\rm Ma}_{\rm turb}^2 \equiv v_{\rm turb}^2/c_{\rm s}^2 \sim \alpha$, for eddies with a dominant length scale $\ell \sim \sqrt{\alpha} H_{\rm g}$ and turnover time $t_{\rm eddy} \sim \varOmega^{-1}$.  Both the pressure reversal and particle stopping (applied to $\taus \lesssim 1$ since larger solids will fly through these eddies) criteria give the same requirement for trapping: $\alpha \gtrsim (H_{\rm g}/R)^2 \sim 10^{-3}$, a fairly standard value.

The effectiveness depends on details beyond these simple estimates.  Inverse cascades to long-lived, shear-elongated structures will enhance trapping.  Some MRI simulations have found strong clumping \citep{jkh06,nature07}, and work is ongoing to understand the robustness of these results.

\subsubsection{Vortex Trapping}
The possibility of particle trapping in long-lived vortices receives considerable attention, at least in part due to the beautiful analogy with Jupiter's red spot.  We ignore the relevant question of vortex stability \citep[see][]{bm05}, and derive two results: (1) the trapping of particles in anti-cylonic vortices and (2) a constraint on vortex size in the (realistic) case that the gas rotation is sub-Keplerian.  The latter result may be new.  I will not discuss concentration by vortices at the dissipation scale of Komogorov turbulence  \citep{cz01}, since the time and spatial scales may be too short to be relevant \citep{chs08}.

Consider a region of constant $z$-vorticity; $\omega = \partial_x U_y - \partial_y U_x$, where $x$ and $y$ are the local radial and azimuthal coordinates in a frame rotating uniformly with the vortex center. The velocity field in the vortex must satisfy
\begin{equation}\label{eq:Saffvort}
U_x = \epsilon \varOmega_{\rm V} y  
~,~U_y = -{\varOmega_{\rm V} \over \epsilon} x\, ,
\end{equation}
where $\epsilon$ is the $x$:$y$ axis ratio, and $\varOmega_{\rm V}$ is the frequency of fluid oscillations.   The vorticity is $\omega = -\varOmega_{\rm V}(1+\epsilon^2)/\epsilon$, but the  the excess vorticity $\omega_{\rm E} = \omega - \omega_{\rm K} = \omega + 3 \varOmega/2$ is the more relevant quantity in (here Keplerian) shear flows.

We mention two specific vortex solutions, but will keep our derivations general.  The \citet[and references therein]{saf92} vortex solution with $\varOmega_{\rm V} = (3/2)\varOmega \epsilon/(1- \epsilon)$ matches onto the Kepler shear flow $U_{\rm y} = -(3/2)\varOmega x$ for infinite azimuthal elongation: $\epsilon \rightarrow 0$.  
The ``planet" solution of \citet{gng87} with $\varOmega_{\rm V} = \epsilon \varOmega \sqrt{3/(1-\epsilon^2)}$ exactly solves the compressible continuity equation.   

To find the particle trapping criterion we start with the pressure forces in the vortex. Balancing against tidal gravity, Coriolis and inertial accelerations gives
\begin{subequations} 
\begin{eqnarray} 
{1 \over \rho_{\rm g}}{\partial P \over \partial x} &=& 3 \varOmega^2 x + 2 \Omega U_y -U_y{\partial U_x \over \partial y} = \left(3 \varOmega^2 - {2 \varOmega \varOmega_{\rm V} \over \epsilon} + \varOmega_{\rm V}^2\right)x \, ,\\ 
{1 \over \rho_{\rm g}}{\partial P \over \partial y} &=& - 2 \Omega U_x -U_x{\partial U_y \over \partial x} =  \left(- {2 \epsilon \varOmega \varOmega_{\rm V}} + \varOmega_{\rm V}^2\right)y\, . 
\end{eqnarray} 
\end{subequations} 
The divergence of the particle velocity, $\vc{v}_{\rm p}$, then follows from equation (\ref{eq:driftrule})
\begin{equation} \label{eq:divvort}
\nabla\cdot \vc{v}_{\rm p} = \nabla \cdot \Delta \vc{v} = {\ts \over \rho} \nabla^2 P = 2 \taus \left( \omega_{\rm E} + \varOmega_{\rm V}^2/\varOmega \right)\, .
\end{equation} 
Thus particle concentration requires $\omega_{\rm E} < -\varOmega_{\rm V}^2/\varOmega < 0$, i.e. anticylonic relative vorticity.  
 Our result only holds for $\taus \ll 1$, but even for the loose coupling only anticylonic vortices trap particles.  See \citet{chav00} for details.

For vortices embedded in sub-Keplerian gas disks,\footnote{Local vortex solutions are unaffected by a constant global pressure gradient, except for a Galilean velocity transform by $-\eta v_{\rm K}$.} the centers are not stationary points for particle trapping.  No drag forces would be exerted there to enforce a sub-Keplerian particle orbit.   The answer -- that the stationary point remains radially centered on the vortex, but moves forward in azimuth -- is obvious in retrospect.  In the Keplerian frame, particles trapped in a sub-Keplerian vortex experience a radial Coriolis force, $f_{{\rm Cor},x} = -2\varOmega \eta v_{\rm K}$.   This  must be balanced by the radial drag force of the vortex flowing over the stationary particle as $f_{{\rm Cor},x} + U_x(y_s)/\ts = 0$.   The stationary point is
\begin{equation} \label{eq:ystat}
y_s =  2 {\taus \eta v_{\rm K} \over \epsilon \varOmega_{\rm V} }\approx {4 \over 3}{\taus \eta R \over \epsilon^2}\, ,
\end{equation} 
and the final approximate expression is for an elongated ($\epsilon \ll 1$) Saffman vortex (i.e.\ a weak perturbation to the Keplerian vorticity).  The constraint that $y_s$ lie inside the vortex imposes a minimum size for particle trapping, which becomes more stringent for longer stopping times.

\subsubsection{Drag Instabilities}\label{s:inst}
A surprising consequence of drag forces in disks is that they drive instabilities which clump particles.  This possibility was first explored by \citet[hereafter GP00]{gp00}.  Instead of calculating microscopic drag forces as in equation (\ref{eq:Fdrag}), they developed a single fluid model for a (height-integrated) particle layer subject to turbulent boundary layer drag.  They found a robust clumping instability for various formulations of turbulent momentum transport.  The crucial factor -- which they demonstrate with an intuitive ``toy model" -- is that the frictional drag acceleration must vary with the surface density of the particle layer.

The streaming instability (SI) of \citet{yg05} is also quite robust and leads to clumping, but the setup of the model is different.  The calculation is a bit more detailed, including the 3D coupled motions of both particles (a pressureless fluid in the analytic treatment) and gas.  However the instability arises spontaneously from a laminar background, and the required physical ingredients are minimal: (1) local Keplerian dynamics (axisymmetric and unstratified in the simplest form), (2) drag forces which act on both particles and (by Newton's Third Law) gas, and (3) a global pressure gradient.  The microscopic drag acceleration on a particle, $\vc{f}_{\rm d,p} = {\bf F}_{\rm d}/m_{\rm p} = -\Delta \vc{v}/\ts$ is independent of the spatial mass density of particles, $\rho_{\rm p}$, unlike the turbulent boundary layer drag of GP00.  Instead it is the back reaction on the gas, $\vc{f}_{\rm d, g} = -(\rho_{\rm p}/\rho_{\rm g})\vc{f}_{\rm d,p} = (\rho_{\rm p}/\rho_{\rm g})\Delta \vc{v}/\ts$, which allows particle clumping to mediate the instability.  Linear perturbations to $\rho_{\rm p}$ interact with the background aerodynamic drift, $\Delta \vc{v}$, which was described in \S\ref{s:drift}.

The ultimate energy source for all drag instabilities -- the secular instability of GP00, SI and KHI --  is pressure-supported, sub-Keplerian gas motion.   
While the above discussion emphasizes robustness, drag forces do not replace the need for an angular momentum transport mechanism in accretion disks (MRI turbulence is currently the leading candidate).  Drag affects the gas only where mass loading is significant, $\rho_{\rm p} \gtrsim \rho_{\rm g}$, i.e.\ in the sedimented midplane.  Also drag effects are strongest for moderate coupling (near $\tau_{\rm s} \sim 1$).  Instabilities are weak for small particles ``slaved" to the gas, or for large bodies that mostly ignore the gas.  Finally, at least for SI, angular momentum transport is inwards, the wrong direction for stellar accretion \citep{yj07,jy07}.
However, as discussed in \S\ref{s:turbGI}, drag instabilities play a crucial role in planetesimal formation theories, especially if clumping leads to gravitational collapse.

\section{Concluding Remarks}\label{s:concl}
Planetesimal formation remains unsolved, despite considerable progress on several fronts.  Laboratory studies of collisions place constraints on sticking assumptions \citep{bw08}.  Numerical simulations coupled 
with analytic theory provide insights to physical processes in disks.  The response of particles to -- and the effect of particles on -- turbulent gas is an especially promising area.  Contrary to previous expectations, simulations show that gravitational collapse of $\gtrsim 10$ cm solids into planetesimals occurs in  turbulent protoplanetary disks \citep{nature07}.   Forming planetesimals directly by gravitational collapse from chondrule-sized, $\lesssim 1$ mm, solids would be an elegant solution.  But such small particles are more tightly coupled to the gas, making it difficult to understand (analytically or numerically) how they might collapse in standard disk models.

Sadly numerical (and even laboratory) experiments lack direct observational confirmation.  But the initial conditions and basic ingredients of theoretical models can be compared to the meteoritic record and observations of protoplanetary disks.  Spitzer is delivering exciting, detailed results.  Soon Herschel and ALMA will further expand our knowledge of the conditions in -- and structure of -- disks.


\end{document}